\documentclass[aps,prl,twocolumn,superscriptaddress,showpacs]{revtex4-1}
\usepackage[latin1,utf8]{inputenc}
\usepackage[pdftex]{epsfig}
\usepackage{graphicx}
\usepackage{color}
\usepackage{latexsym}
\usepackage[pdftex,colorlinks=false]{hyperref}
\usepackage[dvipsnames]{xcolor}
\hypersetup{colorlinks,
                 citecolor=blue,
                 filecolor=blue,
                 linkcolor=black,
                 urlcolor=blue,
                 pdftex}
\begin{document}
\title{Magnetic Excitations in the Ground State of $\mathrm{Yb_2Ti_2O_7}$}

\author{Viviane Pe\c canha-Antonio}
\email{v.pecanha.antonio@fz-juelich.de}
\affiliation{J\"ulich Centre for Neutron Science (JCNS) at Heinz Maier-Leibnitz Zentrum (MLZ), Forschungszentrum J\"ulich GmbH, Lichtenbergstr. 1, D-85747 Garching, Germany}
\affiliation{Physik-Department, Technische Universit\"at M\"unchen, D-85747 Garching, Germany}

\author{Erxi Feng}
\affiliation{J\"ulich Centre for Neutron Science (JCNS) at Heinz Maier-Leibnitz Zentrum (MLZ), Forschungszentrum J\"ulich GmbH, Lichtenbergstr. 1, D-85747 Garching, Germany}

\author{Yixi Su}
\email[corresponding author:]{y.su@fz-juelich.de}
\affiliation{J\"ulich Centre for Neutron Science (JCNS) at Heinz Maier-Leibnitz Zentrum (MLZ), Forschungszentrum J\"ulich GmbH, Lichtenbergstr. 1, D-85747 Garching, Germany}

\author{Vladimir Pomjakushin}
\affiliation{Laboratory for Neutron Scattering and Imaging (LNS), Paul Scherrer Institute, Villigen CH-5232, Switzerland}

\author{Franz Demmel}
\affiliation{ISIS Facility, Rutherford Appleton Laboratory, Chilton, Didcot OX11 0QX, United Kingdom}

\author{Lieh-Jeng Chang}
\affiliation{Department of Physics, National Cheng Kung University, Tainan 70101, Taiwan}

\author{Robert J. Aldus}
\affiliation{J\"ulich Centre for Neutron Science (JCNS) at Heinz Maier-Leibnitz Zentrum (MLZ), Forschungszentrum J\"ulich GmbH, Lichtenbergstr. 1, D-85747 Garching, Germany}

\author{Yinguo Xiao}
\affiliation{J\"ulich Centre for Neutron Science (JCNS) and Peter Gr\"unberg Institut (PGI), Forschungszentrum J\"ulich GmbH, D-52425 J\"ulich, Germany}

\author{Martin R. Lees}
\affiliation{Department of Physics, University of Warwick, Coventry CV4 7AL, United Kingdom}

\author{Thomas Br\"uckel}
\affiliation{J\"ulich Centre for Neutron Science (JCNS) and Peter Gr\"unberg Institut (PGI), Forschungszentrum J\"ulich GmbH, D-52425 J\"ulich, Germany}
\affiliation{J\"ulich Centre for Neutron Science (JCNS) at Heinz Maier-Leibnitz Zentrum (MLZ), Forschungszentrum J\"ulich GmbH, Lichtenbergstr. 1, D-85747 Garching, Germany}

\begin{abstract}
We report an extensive study on the zero field ground state of a powder sample of the pyrochlore $\mathrm{Yb_2Ti_2O_7}$. A sharp heat capacity anomaly that labels a low temperature phase transition in this material is observed at 280 mK. Neutron diffraction shows that a \emph{quasi-collinear} ferromagnetic order develops below $T_\mathrm{c}$ with a magnetic moment of $0.87(2)\mu_\mathrm{B}$. High resolution inelastic neutron scattering measurements show, below the phase transition temperature, sharp gapped low-lying magnetic excitations coexisting with a remnant quasielastic contribution likely associated with persistent spin fluctuations. Moreover, a broad inelastic continuum of excitations at $\sim0.6$ meV is observed from the lowest measured temperature up to at least 2.5 K. At 10 K, the continuum has vanished and a broad quasielastic conventional paramagnetic scattering takes place at the observed energy range. Finally, we show that the exchange parameters obtained within the framework of linear spin-wave theory do not accurately describe the observed zero field inelastic neutron scattering data.
\end{abstract}
   
\maketitle
\section{Introduction}

A system is said to be frustrated when the energies of all the competing interactions driving its spin-spin correlations cannot be simultaneously minimised. Magnetic frustration is, then, the mechanism that may inhibit conventional long-range order down to $T=0$ K. The pyrochlore family of chemical formula $A_2B_2\mathrm{O_7}$, where $A$ is a trivalent rare earth ion and $B$ is a tetravalent transition metal, is notable for presenting a wide range of exotic magnetic phenomena at low temperatures \cite{RevModPhys.82.53}. The reason for that is encoded in the geometry of the underlaying crystallographic lattice, which forms interpenetrating corner sharing tetrahedra in the cubic space group ${F d \bar{3} m}$ and, hence, provides the perfect architecture for geometric frustration.  

As a possible candidate for quantum spin ice (QSI), where emergent $U$(1) gauge field and exotic excitations are expected \cite{0034-4885-77-5-056501}, $\mathrm{Yb_2Ti_2O_7}$ has attracted intense research in recent years \cite{PhysRevLett.106.187202,PhysRevX.1.021002,Lieh,PhysRevLett.119.057203}. The crystal field acting on the magnetic $^{2}F_{7/2}$ Yb$^{3+}$ leads to four well separated Kramers doublets in which the gap between the predominantly $m_J=\pm1/2$ ground state and the first excited state is around 700 K \cite{0953-8984-13-41-318,PhysRevB.92.134420}. This renders $\mathrm{Yb_2Ti_2O_7}$ an authentic effective $S=1/2$ spin system, particularly suited for theoretical investigations on the QSI physics based on an anisotropic spin Hamiltonian model \cite{PhysRevLett.106.187202,PhysRevX.1.021002,PhysRevLett.119.057203,PhysRevLett.109.097205}.

The strong sample dependence exhibited by the magnetic behavior of $\mathrm{Yb_2Ti_2O_7}$ at low temperatures is also particularly noteworthy. One suggestion for the discrepancies is the so-called ``stuffing'', in this sense meaning that the Yb$^{3+}$ ions expected to occupy the $A$ lattice site are randomly distributed on the $B$ site, which in principle should host only Ti$^{4+}$ ions. This substitution was shown to occur on $2.3$\% of the $B$ sites in the sample reported in the studies presented in Refs. \cite{PhysRevB.86.174424,PhysRevB.93.064406}, a rather small amount of site disorder that can suppress the first order phase transition at $T_\mathrm{c}$ \cite{PhysRevB.95.094407}. Recently, however, two powder samples claimed to be stoichiometric show different long-range magnetic orders evolving at low temperatures. Ref. \cite{PhysRevB.93.064406} proposes an ice-like splayed ferromagnetic ground state, in which the magnetic moments are canted $(14 \pm 5)^{\circ}$ from the cubic $\langle$100$\rangle$ axis, in a two-in-two-out arrangement. Ref. \cite{0953-8984-28-42-426002} proposes an all-in-all-out ferromagnetic structure, considering that the representation analysis within the ${F d \bar{3} m}$ structural space group fails in reproducing the magnetic structure observed in their sample below $T_\mathrm{c}$.

Controversies also exist in the reported spin dynamics of the ground state of $\mathrm{Yb_2Ti_2O_7}$. Linear spin-wave theory could successfully reproduce the high-field propagating spin-waves present at low temperatures in the material \cite{PhysRevX.1.021002,PhysRevLett.119.057203}. On the other hand, to date, the predicted gapped magnetic excitations in zero field, consequence of the exchange anisotropy in the Hamiltonian, have not been confirmed experimentally. The reported spin dynamics displays little or no change while increasing the temperature from 50 mK to at least 2 K in both single crystal and powder samples \cite{PhysRevB.93.064406,PhysRevB.92.064425}. 

The applicability of linear spin-wave theory is possibly limited by the quantum fluctuations exhibited in the compound at low fields. The mechanism giving rise to these fluctuations is still debated. It has been suggested that the proximity of $\mathrm{Yb_2Ti_2O_7}$ to a phase boundary between two competing magnetic ground states is a possible cause for the dynamical character of the ground state \cite{PhysRevLett.115.267208,PhysRevB.95.094422}. Ref. \cite{PhysRevLett.115.267208} reports that an order-by-disorder mechanism may drive the system to a XY $\Psi_2$/$\Psi_3$ magnetic phase before it eventually adopts an energetically favourable splayed ferromagnetic state, in a \emph{double phase transition} process. Indeed, a new set of exchange parameters place the compound very close to the boundary between the $\Psi_3$ and the splayed ferromagnetic phases in a calculated semiclassical phase diagram (see Supplemental Material of Ref. \cite{PhysRevLett.119.057203}). The proximity to a critical point would in principle explain the sensitivity of $\mathrm{Yb_2Ti_2O_7}$ to weak disorder, as is the case in stuffed samples. 

In this work we present a complete study of a $\mathrm{Yb_2Ti_2O_7}$ powder sample. We report heat capacity, neutron diffraction and high-resolution inelastic neutron scattering measurements. Heat capacity at constant pressure displays a sharp phase transition at $T_\mathrm{c}=280$ mK. A second, higher temperature broad hump is also observed at 2.5 K. Neutron diffraction confirms the development of a \emph{quasi-collinear} ferromagnetic long-range order below $T_\mathrm{c}$ with a strongly reduced ordered moment of the Yb$^{3+}$ ions. High resolution neutron spectroscopy displays the evolution, upon cooling, of the magnetic correlations in the system. From about 2.5 K, the onset of ferromagnetic correlations is suggested by a broad diffuse quasielastic scattering prominent at low-$|Q|$. Additionally, an inelastic continuum of scattering develops at energy transfers around 0.6 meV. These features exhibit progressively stronger magnitude when the temperature is decreased down to 500 mK. At 50 mK, an energy gap of 0.17 meV opens to form a flat mode in the spectra. An additional sharp feature is visible at 0.1 meV. The significant amount of quasielastic scattering remaining at this temperature once again shows that, despite the long-range order, the magnetic moment of the Yb$^{3+}$ is not fully ordered but fluctuates down to temperatures as low as 50 mK. Using three different sets of exchange parameters, reported in \cite {PhysRevX.1.021002}, \cite{PhysRevLett.119.057203} and \cite{PhysRevB.92.064425}, we calculate the powder-averaged spectra of the spin wave excitations expected to emerge in the ground state and perform a qualitative comparison between the predictions and our experimental results in $\mathrm{Yb_2Ti_2O_7}$. 

\section{Experimental details}

The powder sample was prepared using the standard solid state reaction method. Stoichiometric quantities of $\mathrm{Yb_2O_3}$ and $\mathrm{TiO_2}$ oxides were mixed, pressed into pellets and sintered at 1300 $^{\circ}$C for several days with intermediate grindings \cite{PhysRevB.89.224419}.

High resolution neutron powder diffraction was performed at the HRPT diffractometer at the Paul Scherer Institute (PSI) \cite{S092145269901399X}, Switzerland, at 2 and 150 K. The powder was loaded in a vanadium sample holder and cooled in a liquid He cryostat. A neutron wavelength of 1.155 {\AA} was chosen in order to have an optimal combination of intensity and resolution.

The molar heat capacity at constant pressure ($C_\mathrm{p}$) of our $\mathrm{Yb_2Ti_2O_7}$ powder was measured using a Physical Property Measurement System (PPMS) from Quantum Design. After the addenda measurement, a pressed powder sample was mounted on a dilution refrigerator insert. The size and shape of the sample were chosen in order to optimize the thermal coupling between sample and puck, which was maintained at more than 70\% for all temperature points measured above 100 mK.

Polarized and unpolarized neutron diffraction measurements were carried out at DNS at the Heinz Maier-Leibinitz Zentrum (MLZ), Garching, Germany. Approximately 6 g of powder were packed in an annular cylinder sample holder made with oxygen-free copper and sealed in a He atmosphere. Measurements were taken in the temperature range from 100 mK up to 600 mK in a dilution insert installed in a top-loading CCR cryostat. Particular attention was paid to the thermalisation of powders in the mK temperature regime. Typical neutron polarisation rate at the chosen neutron wavelength $\lambda=4.2$ {\AA} is about 96\%. The standard procedures such as flipping-ratio correction and normalisation of detector efficiency have been applied for data analysis.

Finally, high resolution inelastic neutron scattering was performed at the backscattering inverted geometry time-of-flight spectrometer OSIRIS at ISIS with a fixed neutron final energy $E_f=1.845$ meV selected by a pyrolytic graphite PG(002) analyzer. For this measurements, 2 g of the same sample used in the other experiments were loaded in a copper annular can and sealed in a He atmosphere. One day before the first measurement at base temperature, the sample was mounted in a Kelvinox Oxford Instruments dilution insert and cooled down to 50 mK. Data were collected upon warming for at least 12 hours at different temperatures in several runs of approximately two hours each. After individual checks, datasets of the same temperature were combined. The mean energy resolution provided by OSIRIS, as high as $25$ $\mu$eV, can be seen to not change considerably with energy transfer. A dynamic range of -0.2 to 1.2 meV is accessed. After the inelastic scans at 50 mK and 300 mK were finished, additional diffraction data were also collected at OSIRIS for approximately 2 hours. These data offer a considerably better ($\Delta{d}/d\sim10^{-3}$) resolution than our DNS data. 

\section{Results}
\subsection{Sample stoichiometry}

\begin{figure*}
\includegraphics[width=15cm]{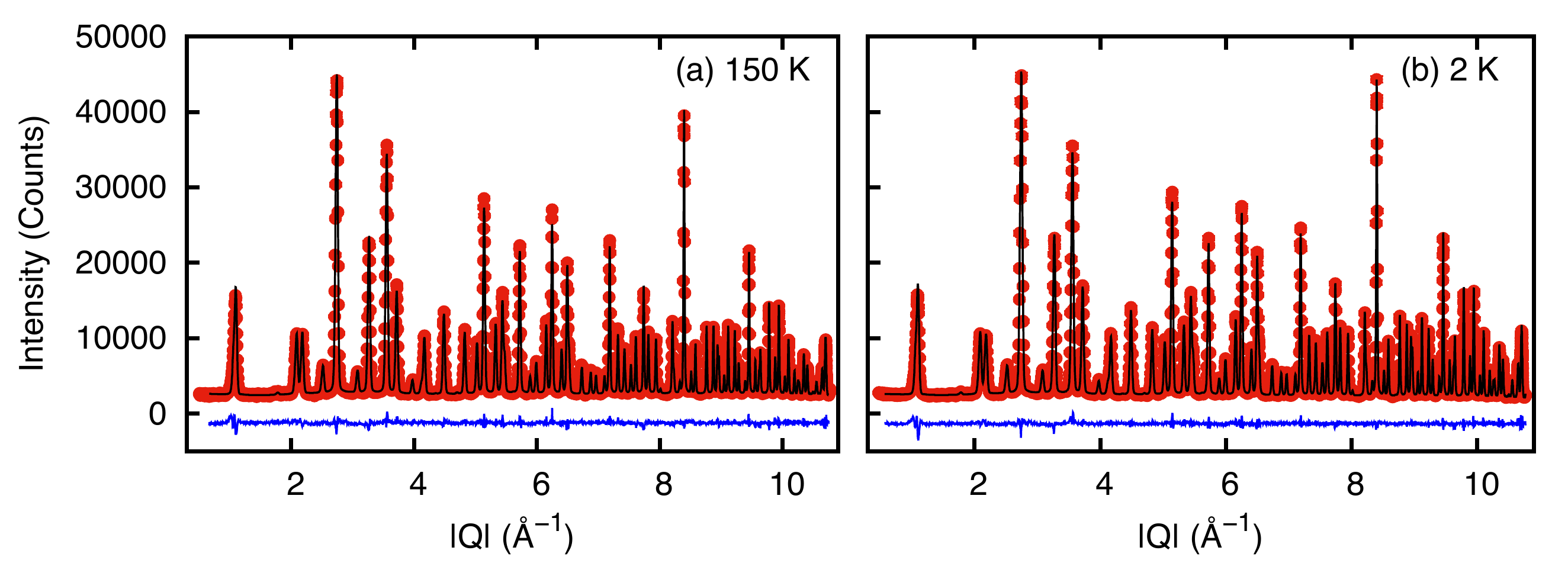}
\caption{High resolution neutron diffraction data (red) and refinement (black) for (a) 150 K and (b) 2 K. The difference between the calculated and the measured intensities is given by the blue line below each pattern.}\label{structure.pdf}
\end{figure*}

\begin{table*}
\begin{tabular}{|c|c|c|c|c|}
\hline
Temperature (K) & Occupancy of Yb$^{3+}$ at 16c site  & Lattice parameter (\AA)&\emph{x}&$R_{wp}$\\
\hline
\hline
\multicolumn{5}{|c|}{Stoichiometric model}\\
\hline
2 &-& 10.01275(2) & 0.33140(5) & 0.0671 \\
150  &-& 10.02061(2) & 0.33105(6) & 0.0787 \\
1.5  (Yaouanc \emph{et al.})&-& 10.0220(5)& 0.332(1)&-\\
150  (Ross \emph{et al.})&-& 10.01111(3) & 0.33122(3)& 0.0414  \\
\hline
\hline
\multicolumn{5}{|c|}{Stuffed model}\\
\hline
2  &  0.0009(1) & 10.01273(2)& 0.33126(5) & 0.0670 \\
150  & 0.0010(1) & 10.02059(2)& 0.33095(6) & 0.0785\\
150  (Ross \emph{et al.}) & 0.002(1) & 10.01111(3) & 0.33121(3) & 0.0415  \\
\hline
\end{tabular}
\caption{Refined lattice parameters and oxygen 48f Wyckoff position for the stoichiometric and stuffed models. For comparison, results from the \emph{sintered powder} of Ross \emph{et al.} \cite{PhysRevB.86.174424} and Yaouanc \emph{et al.} \cite{0953-8984-28-42-426002} are also shown.}\label{tab2}
\end{table*}

\begin{table}
\begin{tabular}{|c|c|c|}
\hline
Ion& IDP at $2$ K (\AA$^2$)& IDP at $150$ K (\AA$^2$) \\
\hline
\hline
\multicolumn{3}{|c|}{Stoichiometric Model}\\
\hline
Yb$^{3+}$ &0.124(4) & 0.347(5) \\
Ti$^{4+}$& 0.21(1) & 0.31(1)  \\
O${_1}^{2-}$ & 0.309(5) & 0.396(6)  \\
O${_2}^{2-}$ & 0.21(1) & 0.26(1)  \\
\hline
\hline
\multicolumn{3}{|c|}{Stuffed Model}\\
\hline
Yb$^{3+}$ & 0.147(4) & 0.37(5) \\
Ti$^{4+}$/Yb$^{3+}$& 0.12(2) & 0.21(2)  \\
O${_1}^{2-}$ & 0.267(6) & 0.348(7)  \\
O${_2}^{2-}$ & 0.21(1) & 0.26(1)  \\
\hline
\end{tabular}
\caption{Isotropic displacement parameters refined at 2 and 150 K for both models considered in this work.}\label{tab3}
\end{table}

\begin{figure}
\includegraphics[width=8.5cm]{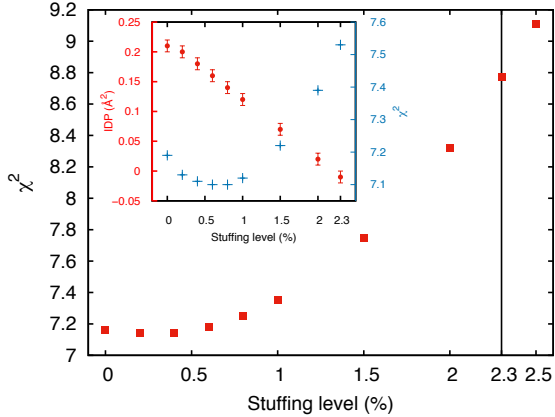}
\caption{The refinement $\chi^2$ plotted versus the stuffing level at 2 K. The minimum values of the uncertainty parameter remain below 1\% stuffing. The solid black line highlights the $\chi^2$ when the 2.3\% stuffing level is reached. The inset shows the IDP and $\chi^2$ versus the stuffing level calculated using FullProf. There is a clear linear correlation between the parameters. As the number of Yb$^{3+}$ ions in the 16c site increases, the IDP value decreases, becoming negative for 2.3\%.}\label{fig2.pdf}
\end{figure}

We performed Rietveld refinements with FullProf \cite{Carvajal199355} on the powder diffraction patterns collected at both 2 and 150 K using HRPT. Initially, the patterns were refined together in order to obtain the best set of instrument-dependent parameters. After that, lattice constants, 48f oxygen free-position ($x$ in the tables) and isotropic displacement parameters (IDP) were refined for each temperature separately \footnote{Actually, given the oxygen environment around the Yb$^{3+}$ site, it is expected that the displacement parameters are anisotropic \cite{SUBRAMANIAN198355}. As the quality of the fitting, in our case, does not respond in a sensible way to the refinement of the anisotropic displacement parameters, we opted to carry out the refinement of the IDP's and analyze their behavior with the stuffing level as discussed in the text.}. The calculated patterns are plotted together with the measured ones in Fig. \ref{structure.pdf}. The refined set of parameters are summarised in Tables \ref{tab2} and \ref{tab3}.  

The oxygen environment of the rare earth ion plays a fundamental role in the crystal-field anisotropy and, consequently, in the splitting of the total angular momentum $J$ ground multiplet of the Yb$^{3+}$ isolated ions. As reported in several studies \cite{RevModPhys.82.53,SUBRAMANIAN198355,Sala,PhysRevB.95.094431}, pyrochlores are prone to oxygen deficiencies, especially at the Wickoff 8b site. A safe estimate of oxygen vacancies is possible because the reflection conditions for the pyrochlore structure allow the Rietveld refinement of the oxygen content unambiguously: that Bragg peaks for which $h+k+l=4n$ and that, additionally, do not match the conditions $h,k,l=4n$ or $h,k,l=4n+2$ (examples are (220),(422),(620), etc.) contain scattering contributions exclusively from O$^{2-}$ ions. The refinement of the occupancy of the 8b position supports a 0.6\% reduced oxygen content in our sample, without an overall change in the other parameters for the stoichiometric model shown in Tables \ref{tab2} and \ref{tab3}.  

Possible stuffing was also considered and a model with Yb$^{3+}$ at the Ti$^{4+}$ site (16c Wickoff position) was refined. The fit parameters are also displayed in Tables \ref{tab2} and \ref{tab3}. Our results for the stuffed model are compared with the results of Ross \emph{et al.} \cite{PhysRevB.86.174424} at 150 K. While refined separately for each temperature, the \emph{stuffing level} shows almost the same value given uncertainties. The lattice parameter and $x$ values also do not change significantly from one model to the other. As widely discussed in Ref. \cite{PhysRevB.86.174424}, the thermal displacement parameters have been shown to be sensitive to stuffing, in particular on the 16c site.

In order to investigate the behavior of the IDP quantitatively, we considered two different scenarios. In the first, we fixed all the parameters obtained in the stoichiometric model. By changing the stuffing level on the 16c Wyckoff position and the oxygen occupancy, which we should modify in order to keep the charge neutrality in the material, we analyzed the variation in the $\chi^{2}$ of the fitting. The result of this analysis can be seen in the main panel of Fig. \ref{fig2.pdf}. The minimum of the $\chi^{2}$ is located between 0 and 1\%. 

In the second scenario, we allow the IDP of the 16c site to vary freely while manually changing the stuffing level and analyzing the behavior of the $\chi^{2}$. All the other parameters were kept fixed. The results are displayed in the inset of Fig. \ref{fig2.pdf}. It can be clearly seen that both IDP and stuffing level are inversely correlated parameters. This can be understood if we note that \emph{(i)} the symmetry and reflection conditions for both Yb and Ti sites are the same and \emph{(ii)} both thermal displacement and ionic substitution may attenuate the measured peak intensities. Consequently, the discrepancy between the fit IDP values presented in Table \ref{tab3} for stuffed and stoichiometric models is in great extent dependent of the differences in scattering cross-sections of Yb (23.4 barns) and Ti (4.35 barns). As the displacement parameters are constrained to the atomic site, it is natural that the calculated IDP values are smaller in the model in which one puts Yb in the place of the less scattering Ti. When we reach 2.3\% stuffing, the fit values for IDP become negative, which is unphysical for a defined quadratic parameter and shows the difficulties in discerning a small stuffing from other general disorder that may be present in a polycrystalline sample using only Rietveld refinement of neutron diffraction data. 

Given the refinement uncertainties and the minimum in the error parameters, we can only establish an upper limit of 1\% to the stuffing level of the $\mathrm{Yb_2Ti_2O_7}$ sample studied here.  

\subsection{Phase Transition and Magnetic Structure}
\begin{figure}
\includegraphics[width=8.5cm]{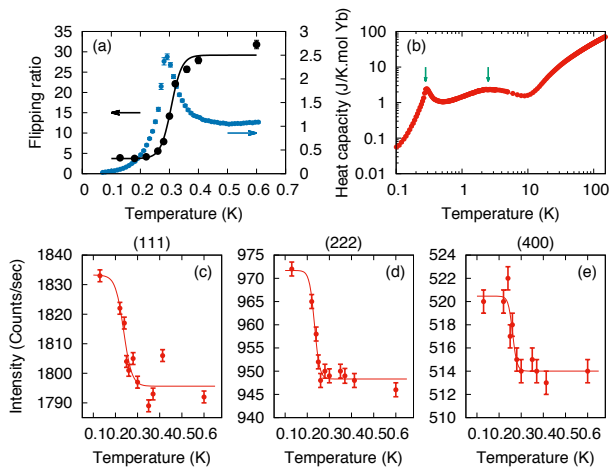}
\caption{(a) The flipping ratio measured at the (111) Bragg position (black). The line is a guide to the eye. The phase transition (blue) in heat capacity is also plotted for comparison. (b) Heat capacity over an extended temperature range. Note the position of the two anomalies (green arrows) at $T_\mathrm{c}=280$ mK and the second hump at $T\sim2.5$ K. (c)-(e) Changes in the intensity of the (111), (222) and (400) Brillouin zone center positions around $T_\mathrm{c}$, measured at DNS.}\label{fig3.pdf}
\end{figure}

Figure \ref{fig3.pdf}(a) shows the sharp heat capacity anomaly exhibited in our sample at $T_\mathrm{c} = 280$ mK. The temperature and amplitude of the $\lambda$-shaped peak are strongly sample dependent. For those single crystals in which the anomaly is seen \cite{Lieh,PhysRevLett.119.057203,PhysRevB.95.094407}, the critical temperature ranges from 160 to 270 mK. For powders, this spread is smaller and temperatures from 210 mK \cite{science} to the 280 mK measured in our sample are reported. 

We used polarized neutrons at DNS to measure the spin flipping ratio of the scattered neutrons in the vicinity of $T_\mathrm{c}$. The results are plotted along with the measured heat capacity in Fig. \ref{fig3.pdf}(a). A drop in the flipping ratio immediately allows one to correlate the heat capacity anomaly with a change in the magnetic structure of the sample. The beam depolarisation suggests the development of ferromagnetic domains below $T_\mathrm{c}$. To conclusively prove that hypothesis, diffraction data were collected using unpolarized neutrons also at DNS. As expected, magnetic Bragg peaks corresponding to a propagation vector $\vec{q}=0$ structure are observed below the phase transition temperature. In Figs. \ref{fig3.pdf}(c)-(e), the changes in intensity of the (111), (222) and (400) Brillouin zone centers around $T_\mathrm{c}$ are shown. This is a clear evidence that the phase transition does correspond to a spontaneous symmetry breaking, in disagreement with what was reported for other powder samples \cite{PhysRevLett.88.077204,PhysRevB.93.064406,PhysRevB.70.180404}.

\begin{figure}
\includegraphics[width=8.5cm]{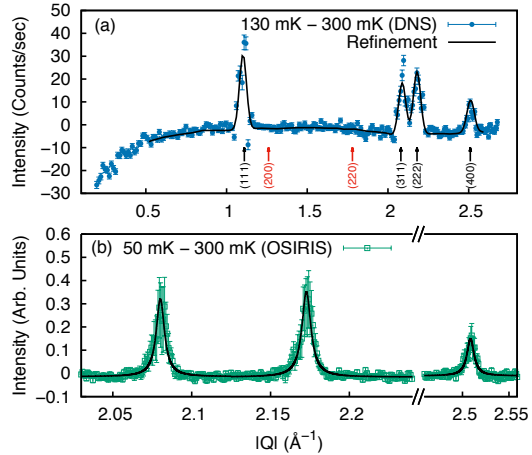}
\caption{(a) Difference between the DNS unpolarized neutron diffraction data collected at 130 and 300 mK (blue points with error bars), interpreted as pure magnetic scattering, along with the refined magnetic structure (black line). The positions of the (200) and (220) magnetic Bragg peaks, present in a splayed ice-like ferromagnet and not apparent in our data, are marked in red. (b) Difference between the OSIRIS diffraction data obtained at 50 and 300 mK. The Lorentzian fit (black line) is used to estimate a minimum correlation length of 780 {\AA} in the ground-state ferromagnetic order.}\label{Diff_resume.pdf}
\end{figure}

The diffraction pattern obtained by subtracting the DNS data measured at 300 mK from the data measured at 100 mK is displayed in Fig. \ref{Diff_resume.pdf}(a). The representation analysis performed is analogous to that of Gaudet \emph{et al.} \cite{PhysRevB.93.064406} and the irreducible representation used is the same, namely, $\Gamma^{(3)}_{9}$. We point out a crucial difference between our data from that of Ref. \cite{PhysRevB.93.064406}: our result does not allow us to affirm that the canting angle relative to the $\langle$100$\rangle$ cubic axis is different from zero to within error ($\pm1^{\circ}$). The (200) and (220) magnetic Bragg peaks that are apparent in the case of the structure being an ice-like splayed ferromagnet, i.e., having an antiferromagnetic component, are not visible in our data, which indicates that our $\mathrm{Yb_2Ti_2O_7}$ sample develops a \emph{quasi-collinear} ferromagnetic structure below the phase transition. The refined magnetic moment of $0.87(2)\mu_\mathrm{B}$ is in agreement with that reported in \cite{PhysRevB.93.064406}, but strongly reduced from the $\approx1.8\mu_\mathrm{B}$ expected from high-field magnetisation measurements \cite{PhysRevB.89.224419}. No evidence supporting the structure observed by Yaounanc \emph{et al.} \cite{0953-8984-28-42-426002} is found, particularly the strong (220) magnetic Bragg peak reported at their sample ground state.

Also shown in Fig. \ref{Diff_resume.pdf}(b) is the difference between the high-resolution diffraction data collected at 50 and 300 mK at OSIRIS. No refinement was performed for these data, but the results are fully consistent with the ones obtained with the unpolarized neutron diffraction at DNS. The average peak width (FWHM) in the measured $|Q|$ range, obtained based on the fitting of a Lorentzian line shape, is about 0.008 \AA$^{-1}$. This would correspond to a magnetic order with a correlation length of at least 780 \AA, thus stating the long-range order character of the ferromagnetic state in $\mathrm{Yb_2Ti_2O_7}$.

\begin{table}
\begin{tabular}{|c|c|c|c|c|c|c|}
\hline
Site & X& Y & Z & m$^a$($\mu_\mathrm{B}$)& m$^b$($\mu_\mathrm{B}$) & m$^c$($\mu_\mathrm{B}$)\\
\hline
\hline
1& $\frac{1}{2}$ & $\frac{1}{2}$ & $\frac{1}{2}$ & 0.00(2) & 0.00(2) & 0.87(1)\\
2& $\frac{1}{4}$ & $\frac{1}{4}$ & $\frac{1}{2}$ & -0.00(2) & -0.00(2) & 0.87(1)\\
3& $\frac{1}{2}$ & $\frac{1}{4}$ & $\frac{1}{4}$ & -0.00(2) & 0.00(2) & 0.87(1)\\
4& $\frac{1}{4}$ & $\frac{1}{2}$ & $\frac{1}{4}$ & 0.00(2) & -0.00(2) & 0.87(1)\\
\hline
\end{tabular}
\caption{Magnetic moments obtained in the refinement shown in Fig. \ref{Diff_resume.pdf}(a). m$^a$, m$^b$ and m$^c$ are the magnetic moment components along the global \^x, \^y and \^z axis, respectively.}\label{tab5}
\end{table}

\subsection{Inelastic Neutron Scattering}

\begin{figure*}
\includegraphics[width=13cm]{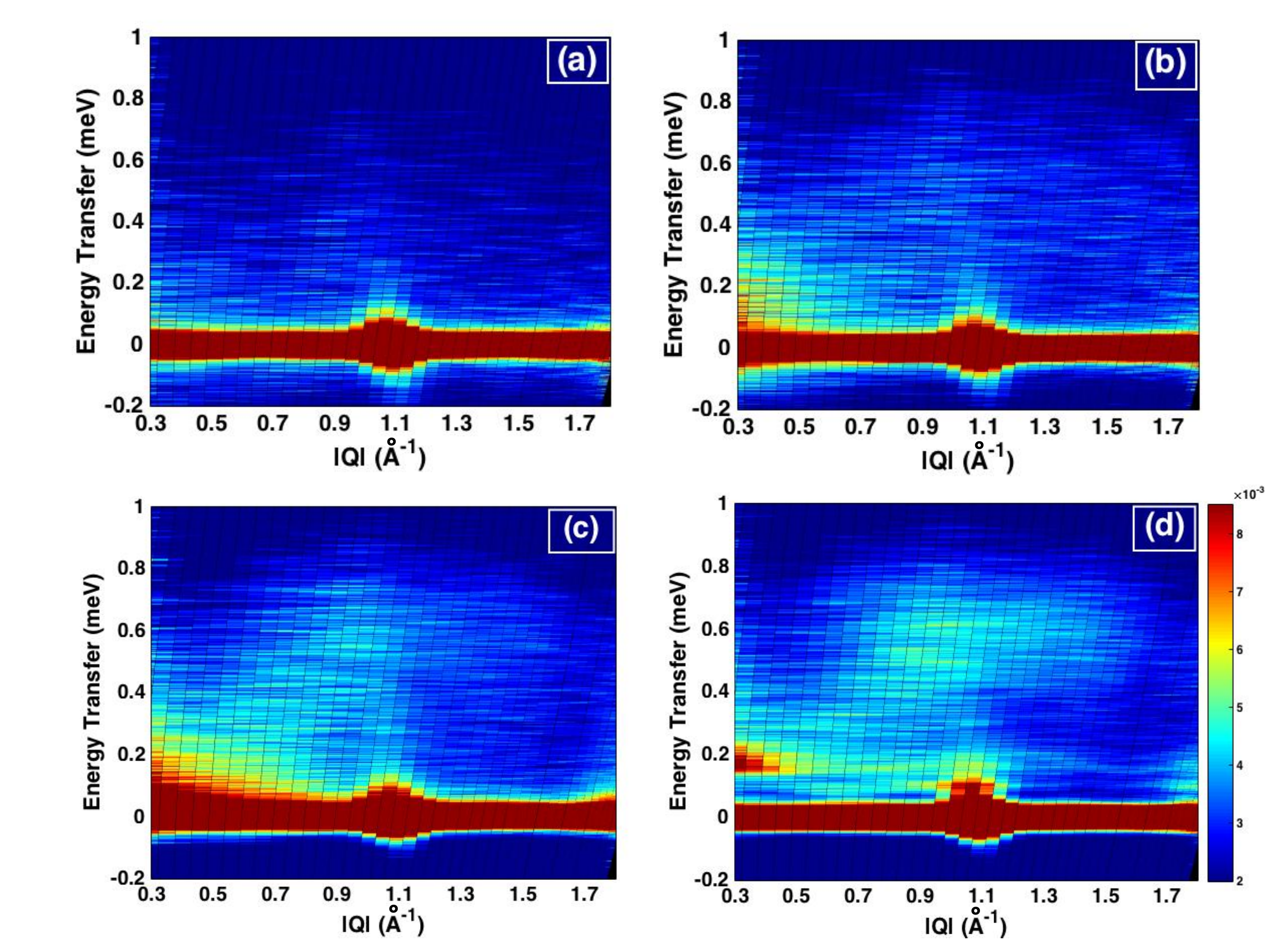}
\caption{Inelastic neutron scattering spectra measured at OSIRIS at (a)10 K, (b) 2.5 K, (c) 500 mK and (d) 50 mK. The momentum transfer interval accessed at ${\hbar}{\omega}=0$ is $|Q|=[0.2,1.8]$ \AA$^{-1}$}\label{colourmaps.pdf}
\end{figure*}

High resolution neutron spectroscopy measurements were carried out in order to investigate the dynamical magnetic behavior of our sample. The contour plots in Fig. \ref{colourmaps.pdf} display the clear response of the measured spectra to the changes in temperature. At 10 K [panel (a)], a typical paramagnetic behavior is exhibited. At 2.5 K [panel (b)], the appearance of a weak quasielastic broad feature at low-$|Q|$ signals the development of dynamical ferromagnetic correlations in the sample. Concomitantly, as will be clearly shown later, a weak diffuse continuum of scattering forms at energy transfers higher than 0.4 meV. Down to 500 mK [panel (c)], both features become stronger and at 50 mK, an energy gap opens at the low momentum transfer region, showing clear features in the spectra at temperatures below $T_\mathrm{c}$.  

\begin{figure}
\includegraphics[width=7.5cm]{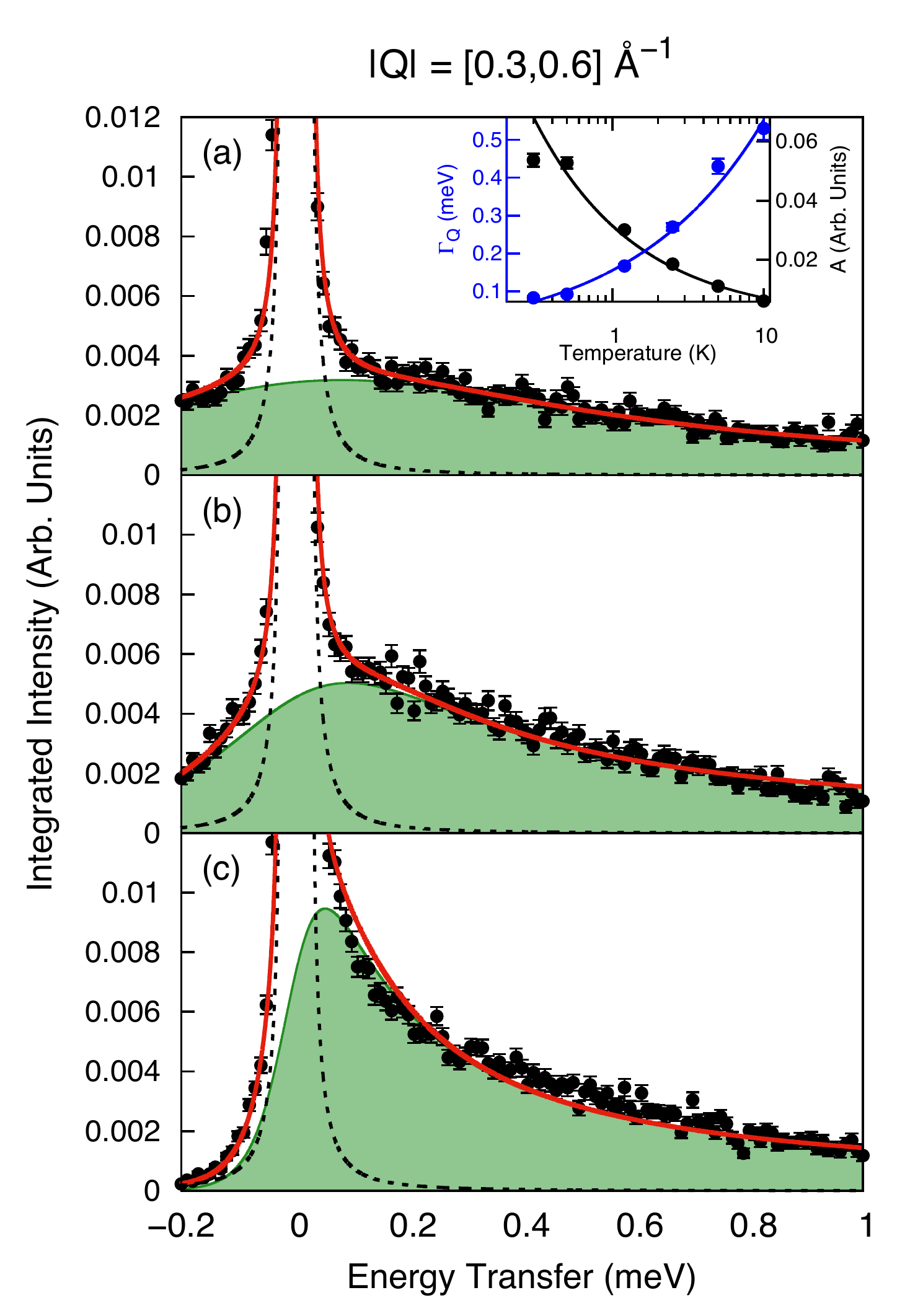}
\caption{Cuts along the energy axis binned in the interval $|Q|=[0.3,0.6]$ \AA$^{-1}$ for (a) 10 K, (b) 2.5 K and (c) 500 mK. The red lines show the sum of all the fit components (see main text) to the measured curves (black points with error bars). The black dashed line is the estimated contribution of the elastic incoherent scattering. The green shaded area corresponds to the fit quasielastic scattering according to Eq. \ref{1}. The inset in (a) shows the temperature evolution of the two fitting parameters $\Gamma_{Q}$ and $A$ for temperatures measured above $T_\mathrm{c}$. The solid lines are guides to the eye.}\label{fig5.pdf}
\end{figure}

\begin{figure}
\includegraphics[width=7.5cm]{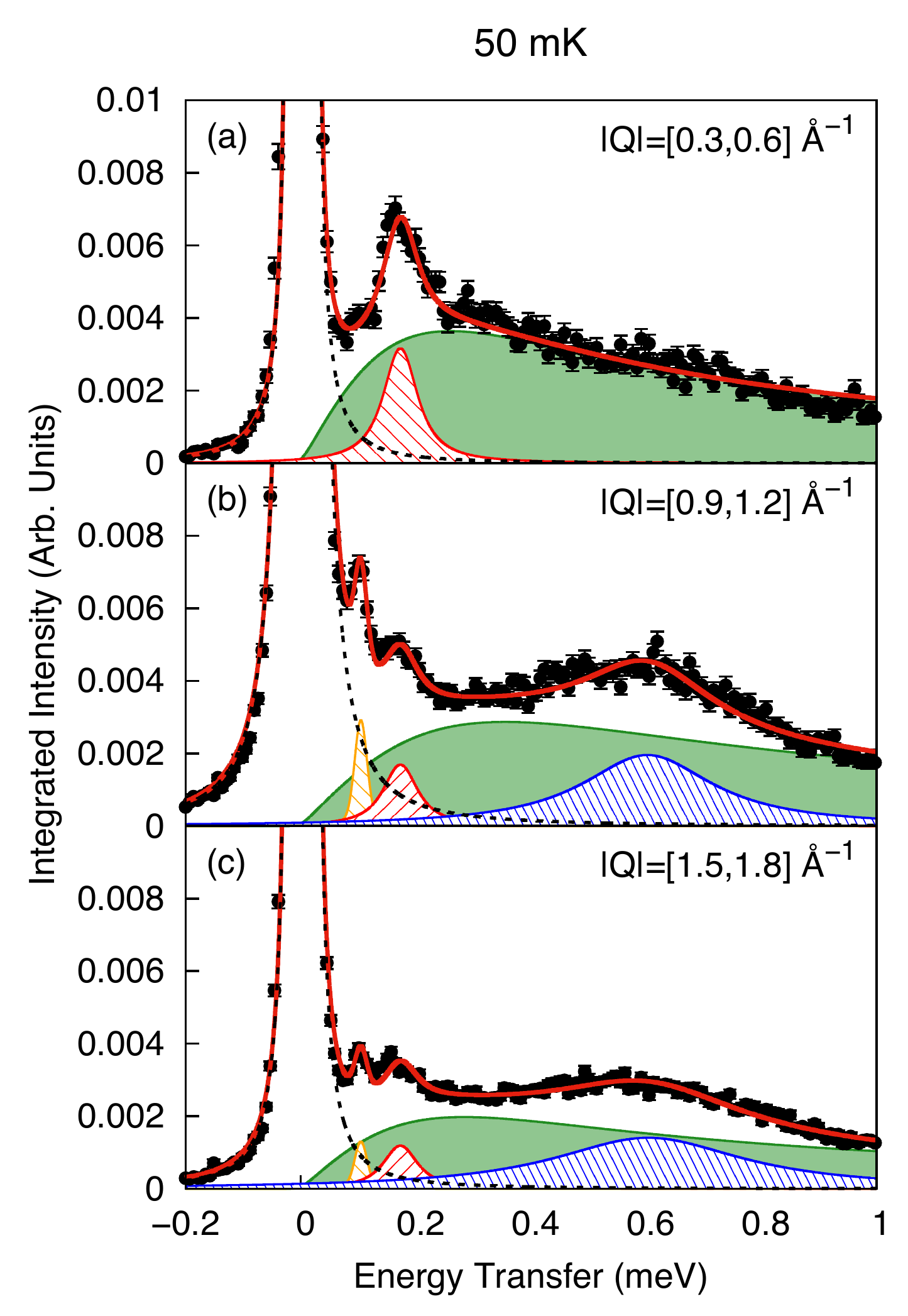}
\caption{Three different cuts at 50 mK for (a) $|Q|=[0.3,0.6]$ \AA$^{-1}$, (b) $|Q|=[0.9,1.2]$ \AA$^{-1}$ and (c) $|Q|=[1.5,1.8]$ \AA$^{-1}$. Note that the (111) Bragg peak is contained in the $|Q|$-range of panel (b) and that it causes the observed broadening of the elastic line. In addition to the estimated quasielastic scattering (green), two sharp magnetic excitations (yellow and red dashed areas, respectively) at ${\hbar\omega}=0.10$ meV (FWHM = 25 $\mu$eV) and at ${\hbar\omega}=0.17$ meV (FWHM = 70 $\mu$eV) and estimated inelastic continuum (blue dashed area) contribute to the observed line shape.}\label{fig5_6.pdf}
\end{figure}

In Figs. \ref{fig5.pdf} and \ref{fig5_6.pdf}, some representative cuts made along the energy axis are shown. The quasielastic contribution to the $|Q|$-cuts is fit using
\begin{equation}\label{1}
I({\hbar}{\omega})=\frac{A}{\pi}\,\frac{{\hbar}{\omega}}{1-\exp({-\beta\hbar\omega})}\frac{\Gamma_{Q}}{(\hbar\omega)^{2}+\Gamma_{Q}^{2}},
\end{equation}
where $\Gamma_{Q}$ is the half width of the quasielastic component, ${\hbar}{\omega}$ is the neutron energy transfer, $\beta={(k_\mathrm{B}T)}^{-1}$, $k_\mathrm{B}$ is the Boltzmann constant and $T$ is the sample temperature. The inelastic (gapped) excitations were fit with a standard Lorentzian convolved with the instrumental resolution function determined from the vanadium scans. 

The evolution of the ferromagnetic correlations, manifested as quasielastic scattering at low-$|Q|$, can be seen in Figs. \ref{fig5.pdf}(a)-(c), which display cuts binned in the range $|Q|=[0.3,0.6]$ \AA$^{-1}$ for different temperatures above $T_\mathrm{c}$. The quasielstic scattering (green shaded area for all the plots) extends over a wide energy range. At lower temperatures, this component narrows and displays an increase in magnitude around zero energy transfer, signaling overall slower spin dynamics [see inset in Fig. \ref{fig5.pdf}(a)]. Closer to the phase transition temperature, at 500 mK [Fig. \ref{fig5.pdf}(c)], this contribution is sharp and asymmetrically broadened at the base of the elastic line. 
  
In Figs. \ref{fig5_6.pdf}(a)-(c) we show cuts for 50 mK binned in the ranges of $|Q|=[0.3,0.6]$ \AA$^{-1}$, $|Q|=[0.9,1.2]$ \AA$^{-1}$ and $|Q|=[1.5,1.8]$ \AA$^{-1}$, respectively. The gapped sharp excitation feature, that is clearly seen for $|Q|< 0.6$ \AA$^{-1}$ in the contour plot of Fig. \ref{colourmaps.pdf}(d), can be seen in all three cuts at ${\hbar\omega}\sim0.17$ meV. This flat-band-like excitation mode seems slightly dispersive, and its intensity is stronger at the magnetic zone center regions around (000) and (111). Despite being outside the measured momentum transfer range, the tail of the dispersions stemming from the (311)/(222) positions can also be observed in the contour plots and cuts of Fig. \ref{fig5_6.pdf}. Its intrinsic width is estimated to be around 70 $\mu$eV based on the fitting shown in Fig. \ref{fig5_6.pdf}(a). Furthermore, in Figs. \ref{fig5_6.pdf}(b)-(c), an additional resolution-limited sharp excitation is present at ${\hbar\omega}\sim0.1$ meV, with the strongest intensity observed near the (111) Bragg position.

Despite the long-range magnetic order and the gapped magnetic excitation modes evident at both 0.17 and 0.1 meV at 50 mK, a broad quasielastic scattering component extending up to at least 1 meV is observed to persist in the whole measured momentum transfer range. This behavior is reported in single crystal samples to manifest in the form of rods of scattering along the $\langle$111$\rangle$ reciprocal lattice directions \cite{PhysRevB.84.174442,PhysRevLett.103.227202,PhysRevB.92.064425,Lieh}. In \cite{PhysRevB.84.174442}, the scattering intensity in the rods grows on approaching $T_\mathrm{c}$ and, at 30 mK, shows reduced spectral weight (depletion) below 0.2 meV in comparison with 500 mK data. In Ref. \cite{PhysRevB.92.064425}, the spectral weight of the rods is little affected when the temperature is increased up to 850 mK. In our case, even though no information about directions in reciprocal space can be accessed, the strong shift of the quasielastic contribution to positive energies ultimately causes the suppression of the scattering magnitude at the energy gap region, thus explaining the depletion observed in Ref. \cite{PhysRevB.84.174442}. 

\begin{figure}
\includegraphics[width=7cm]{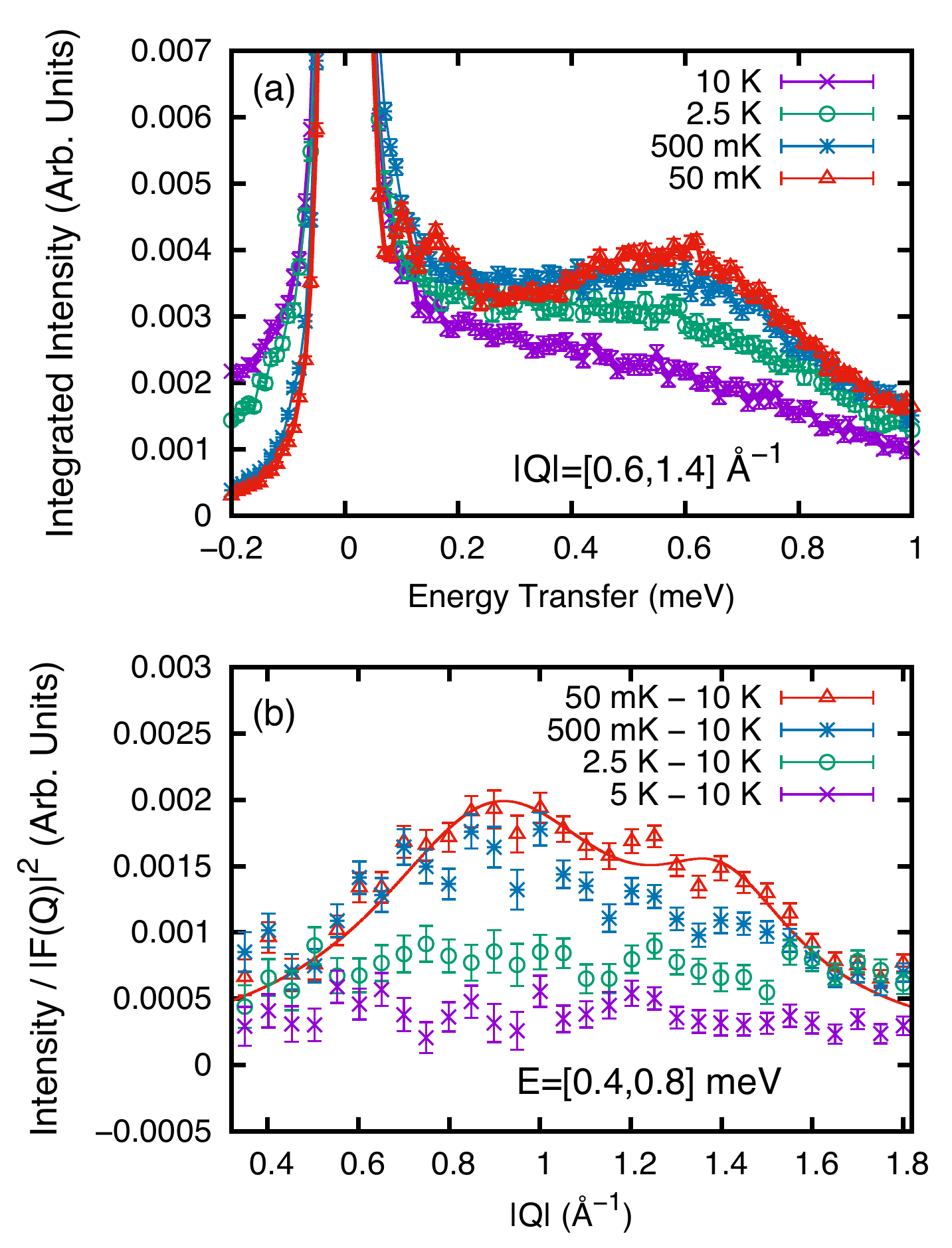}
\caption{(a) A broad cut corresponding to the integrated intensity over the interval $|Q|=[0.6,1.4]$ \AA$^{-1}$ is shown. (b) Energy binned in the range ${\hbar\omega}=[0.4,0.8]$ meV displaying the $|Q|$-dependence of the continuum. The dispersion is double-peaked at 0.9 and 1.4 \AA$^{-1}$. The red line is a guide to the eye.}\label{fig6.pdf}
\end{figure}

A third characteristic feature of the magnetic excitations spectra is the continuum-like inelastic scattering component around 0.6 meV. In Fig. \ref{fig6.pdf}(a), where we plot the temperature dependence of the $|Q|$-cut over a broad momentum transfer range, the scattering intensity in the continuum region of the spectra can be seen to increase steadily relative to 10 K. At 2.5 K the intensity at the tail of the elastic line is approximately constant up to the maximum $\sim 0.6$ meV, after which it starts to decrease. Upon further cooling, this component sharpens to form at 50 mK the prominent peak displayed in the panels of Fig. \ref{fig5_6.pdf} for $|Q|>0.9$ \AA$^{-1}$. 

In Fig. \ref{fig6.pdf}(b), an energy cut integrated over ${\hbar}{\omega}=[0.4,0.8]$ meV is plotted as a function of momentum transfer. The integrated intensity at 10 K was subtracted from the data at each temperature shown and the result was then divided by the magnetic form factor squared of the Yb$^{3+}$ ions. This excitation appears approximately featureless at 5 and 2.5 K , with a slightly increased magnitude for the latter temperature. At 500 mK, a maximum in the dispersion arises $\sim0.9$ \AA$^{-1}$, which is accompanied at 50 mK by another maximum taking place $\sim1.4$ \AA$^{-1}$. Apart from this subtle change at higher-$|Q|$ and unlike the lower energy sharp excitations, the continuum seems to be little affected by the phase transition to the ordered phase.

\begin{figure}
\includegraphics[width=8.5cm]{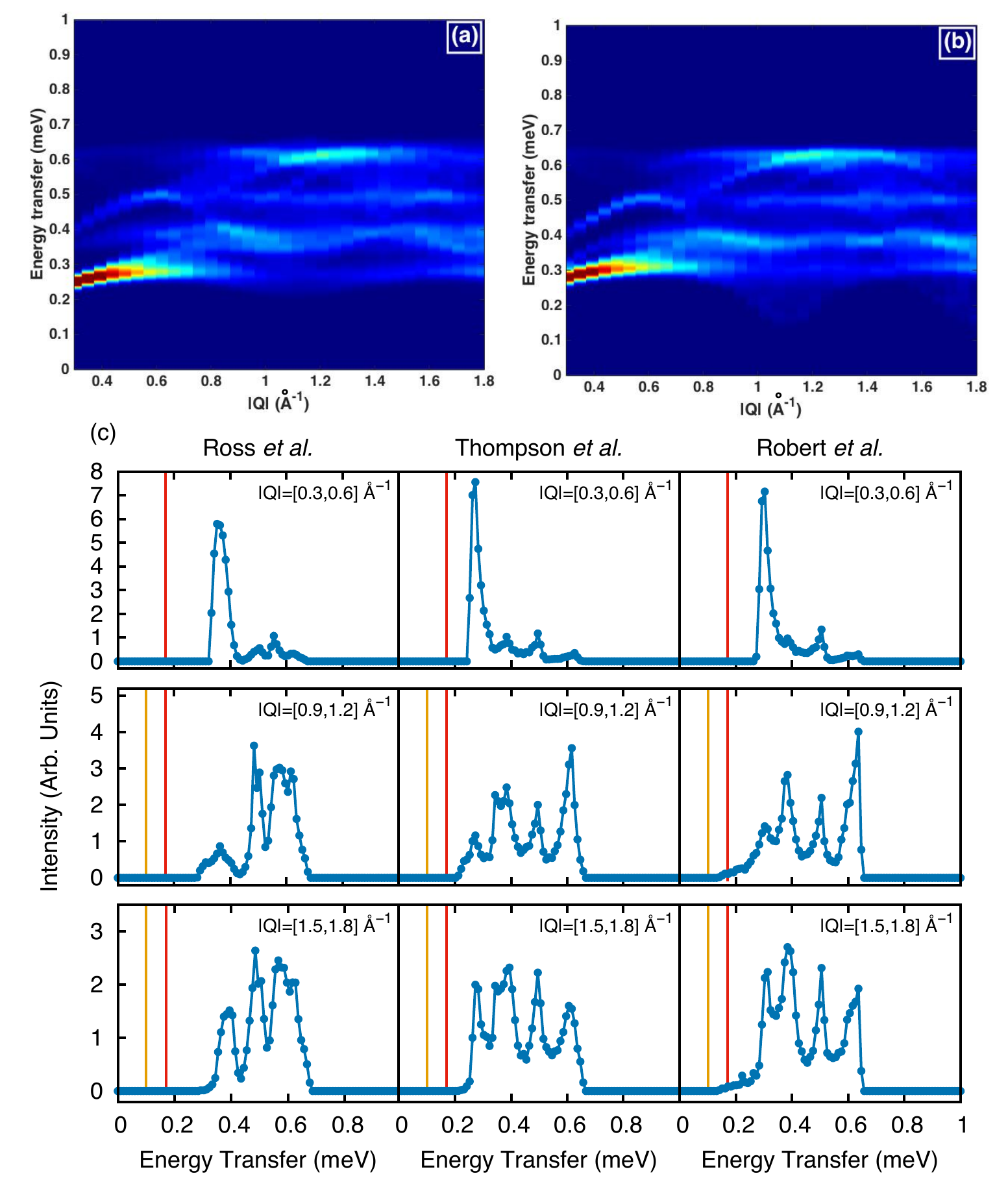}
\caption{Resolution convolved ($25$ $\mu$eV) expected spin-wave spectra calculated using the exchange parameters determined in (a) Ref. \cite{PhysRevLett.119.057203} and (b) Ref. \cite{PhysRevB.92.064425}. (c) Cuts along the energy axes in the same $|Q|$-ranges shown in Fig. \ref{fig5_6.pdf}, including the model of Ref. \cite{PhysRevX.1.021002}. The yellow and red lines mark the position of the two sharp excitations measured in our sample.}\label{colourmaps_2.pdf}
\end{figure}

All the reports of gapped magnetic excitations in $\mathrm{Yb_2Ti_2O_7}$ have been made in applied magnetic fields \cite{PhysRevX.1.021002,PhysRevLett.119.057203}. In single crystals, regardless of sample stoichiometry issues, the field induced order displays clear sharp spin wave branches, which are modeled using linear spin-wave theory. Robert \emph{et al.} \cite{PhysRevB.92.064425} further constrains the parameter space using Monte Carlo and spin dynamics simulations in order to reproduce some features in the diffuse elastic scattering at zero magnetic field. We use the software SPINW \cite{0953-8984-27-16-166002} and the sets of exchange parameters determined in \cite{PhysRevX.1.021002}, \cite{PhysRevB.92.064425} and, more recently, in \cite{PhysRevLett.119.057203}, to calculate the powder averaged, resolution convolved magnetic excitations expected to emerge below $T_\mathrm{c}$ at zero magnetic field. The results are shown in Fig. \ref{colourmaps_2.pdf}. Since the model obtained by Ross \emph{et al.} \cite{PhysRevX.1.021002} was already discussed for powders in Ref. \cite{PhysRevB.93.064406}, we do not reproduce the respective contour plot here. For a quick comparison, the parameters reported in these three works are given in Table \ref{tab6}.

Though successfully reproducing the high field regime of the magnetic excitation spectra, the fit exchange parameters do not give an accurate account of the excitations measured in our sample. The closest resemblance is seen at low-$|Q|$, where a strong dispersive gapped mode is predicted in the powder spectra. However, the branch observed in our experiment is almost flat and the measured gap of $0.17$ meV appears overestimated in all models, but closer to the calculation based on the exchange parameters of Thompson \emph{et al.} No sharp excitation mode $\sim0.1$ meV can be reproduced in the spin-wave calculations based on the exchange parameters of Table \ref{tab6}, as can be clearly seen in the cuts performed in the calculated spectra displayed in Fig. \ref{colourmaps_2.pdf}(c). Given that this mode, of energy of around 1 K, is clearly absent above $T_\mathrm{c}$, we conclude that it can only be associated with the ground state magnetic order. 

\begin{table*}
\begin{tabular}{|c||c|c|c|c||c|c|c|c||c|c|}
\hline
 &$J_{zz}$&$J_{\pm}$&$J_{\pm\pm}$&$J_{z \pm}$&$J_{1}$&$J_{2}$&$J_{3}$&$J_{4}$&$g_{xy}$&$g_{z}$\\
\hline
\hline
Ross \emph{et al.} \cite{PhysRevX.1.021002}& 0.17 & 0.05 & 0.05 & -0.14 & -0.09 & -0.22 & -0.29 & 0.01& 4.32&1.80\\
Thompson \emph{et al.} \cite{PhysRevLett.119.057203}& 0.026 & 0.074 & 0.048 & -0.159 & -0.028 & -0.326& -0.272&0.049&4.17& 2.14 \\
Robert \emph{et al.} \cite{PhysRevB.92.064425}& 0.07 & 0.085 & 0.04 & -0.15 & -0.03 & -0.32& -0.28&0.02&4.09& 2.06\\
\hline
\end{tabular}
\caption{Exchange parameters determined in Refs. \cite{PhysRevX.1.021002,PhysRevLett.119.057203,PhysRevB.92.064425} in both local and global coordinates, with the corresponding Land\`e factors $g_{xy}$ and $g_{z}$. All the values for the exchange parameters are given in meV. }\label{tab6}
\end{table*}

Even thought the four distinct calculated modes are expected to be discernible at higher momentum transfers for a instrumental resolution of $25$ $\mu$eV [see Fig. \ref{colourmaps_2.pdf}(c)], the continuum at higher energies may tentatively be interpreted as a result of the superposition of spin-wave branches caused by powder averaging. However, a similar observation in single crystals \cite{PhysRevB.92.064425,PhysRevLett.119.057203} weakens this hypothesis. We interpret this inelastic excitation as a strong evidence that above the sharp phase transition temperature this compound is not in a conventional paramagnetic state up to at least 2.5 K. In fact, the temperature evolution of the continuum may well be related to the broad hump in heat capacity shown in the Fig. \ref{fig3.pdf}(b). The broad anomaly has an onset at 10 K and a maximum around 2.5 K, temperature below which the heat capacity experiences a smooth decrease down to the sharp phase transition at $T_\mathrm{c}$.
 
Recently, an alternative explanation to the continuum of scattering emerging in $\mathrm{Yb_2Ti_2O_7}$ ground state was suggested. Analyses of terahertz spectroscopy \cite{PArmitage} and inelastic neutron scattering data \cite{PhysRevLett.119.057203} suggest that at low and zero magnetic fields the single and multiple-magnon branches in the compound are expected to overlap. This provides a decay route for the one-magnon excitation and, consequently, suppresses the development of conventional spin-waves in the ordered phase. This effect is know as \emph{quasiparticle breakdown} and it is usually reported in low-dimensional quantum spin systems \cite{Chernyshev,stone}. We point out, however, that the continuum observed here displays significant spectral weight already at 2.5 K $\sim 10T_\mathrm{c}$, much above the effective formation of single-magnon modes and, consequently, above the temperature at which magnon decays can occur.  

Lastly, we point out the striking similarities of the low temperature behavior observed in the isomorphous compound $\mathrm{Yb_2Sn_2O_7}$ and the one presented in our $\mathrm{Yb_2Ti_2O_7}$. The heat capacity of $\mathrm{Yb_2Sn_2O_7}$ displays a sharp anomaly at $T_\mathrm{c}=0.15$ K followed by a broad hump $\sim2$ K \cite{PhysRevLett.110.127207}. In the stannate, the neutron diffraction supports the development of an ice-like splayed ferromagnetic structure below $T_\mathrm{c}$ with a ordered moment of $1.05(2)\mu_\mathrm{B}$. Nevertheless, $\mu$SR measurements show evidence of persistent spin dynamics in the compound ground state \cite{PhysRevLett.110.127207}. Inelastic neutron scattering measurements performed in a different powder sample confirm this scenario. As can be seen in the contour plots of Ref. \cite{PhysRevB.87.134408}, the inelastic continuum is prominent already at 3 K and evolves down to 50 mK, showing weak dispersion and stronger intensity at $|Q|=0.9$ \AA$^{-1}$. No gapped modes are reported in \cite{PhysRevB.87.134408}, what may be a consequence of insufficient instrumental energy resolution (0.1 meV at that experiment). These works show strong evidence that $\mathrm{Yb_2Sn_2O_7}$ may mimic the behavior reported here for $\mathrm{Yb_2Ti_2O_7}$, in spite of the chemical pressure arising from the different occupation of the $B$ site.  

\section{Discussion}

Sample dependence clearly complicates the overall understanding of $\mathrm{Yb_2Ti_2O_7}$ ground state. Once believed to not order down to the lowest temperatures \cite{PhysRevLett.88.077204,PhysRevLett.103.227202}, for many years the sharp anomaly displayed in $\mathrm{Yb_2Ti_2O_7}$ heat capacity could not be attributed to any spontaneous symmetry breaking taking place in the material. Until recently, only the single crystal of Refs. \cite{Lieh,jps_jpsj72_3014} had been shown to display the magnetic Bragg peaks of a ferromagnet below $T_\mathrm{c}$. In the work of Yasui \emph{et al.}, a simple collinear structure with ordered moment of $(1.1\pm0.1)\mu_\mathrm{B}$ is reported. 
 
Later work on powders established the long-range character of the ground state, even though the magnetic structure is still disputed, as pointed out above. The powder of Ref. \cite{PhysRevB.93.064406} displays a very sharp phase transition in $C_\mathrm{p}$ at $T=260$ mK. Neutron diffraction measurements show that the Bragg (111) intensity in their sample is approximately constant while warming from 100 mK to $\sim$350 mK. Ref. \cite{0953-8984-28-42-426002} does not report detailed temperature dependence of the measured intensities, but it is supposed that it is only below the phase transition that the sample develops the all-in-all-out ferromagnetic order. Recent work \cite{PhysRevLett.119.127201}, performed on single crystal, report results in close resemblance to ours. Notwithstanding being performed using the intensities of only three Bragg peaks, the magnetic structure refinement also supports that $\mathrm{Yb_2Ti_2O_7}$ is a canted two-in-two-out ferromagnet with ordered moment of $0.90(3)\mu_\mathrm{B}$ and a canting angle of $8(6)^{\circ}$. 

Our inelastic data is the source of new and surprising behavior. Below the phase transition, the quasielastic scattering is expected to split its spectral weight into elastic (static) contributions and inelastic dynamical scattering in form of coherent spin wave modes. Instead, we observe that at 50 mK inelastic excitations coexist with a quasielastic contribution, which signals spin fluctuations. From these observations, several important questions arise and we discuss some of them below, one by one: 

(i) \emph{Are the fluctuations caused by poor thermalization of the sample at low temperatures?} That is always possible, especially if we consider that the thermalization in powders is inherently poor. However, evidences in our data show that this possibility is improbable. First, the inelastic neutron scattering spectra show clear temperature dependence. Second, the diffraction data collected just after the inelastic scattering measurements above and below the phase transition temperature at OSIRIS, display the Bragg peaks of a long-range ferromagnetic order. In other words, if some small amount of sample remains above $T_\mathrm{c}$, it is unlikely that it would be the single cause for the significant fluctuating component we observe to persist in the $\mathrm{Yb_2Ti_2O_7}$ ground state. 

(ii) \emph{Are the fluctuations caused by sample inhomogeneities?} Arguing against this suggestion, it should be enough to say that the sample displays the phase transition and develops long-range ferromagnetic order, which has been shown to be suppressed in non-stoichiometric samples. We also point out that \emph{all} the inelastic neutron scattering data for $\mathrm{Yb_2Ti_2O_7}$ samples reported to date have shown dynamics in the ground state at zero magnetic field. So, to suggest that the fluctuations are due to sample inhomogeneities would be equivalent to saying that \emph{all} the samples of $\mathrm{Yb_2Ti_2O_7}$ are affected by disorder and do not show the \emph{real} behavior of the compound. 

(iii) \emph{Are the fluctuations caused by an intrinsic dynamical ground state surviving in the long-range ordered regime?} This possibility is the most promising and, given the large body of work showing the eccentricities in the magnetic behavior of $\mathrm{Yb_2Ti_2O_7}$, it is natural to link our results with the models developed for the QSI. Indeed, the ground state excitations we observe are exotic, but a key experimental fact places the compound away from the QSI regime: the ground state static long-range ferromagnetic structure not (totally) disrupted by quantum fluctuations (see, for example, the discussions in \cite{PhysRevB.87.205130} and the recent reviews in \cite{0034-4885-77-5-056501,0034-4885-80-1-016502}). A good \emph{qualitative} description of our observations seems to be achieved when we compare our experimental results with those predicted by the gMFT theory developed in \cite{PhysRevLett.108.037202}. The so-called Coulombic Ferromagnet (CFM) phase, whose elementary excitations are an inelastic continuum of \emph{spinons} instead of conventional spin-waves, displays magnetic order and supports the existence of a gapless photon mode. The settling of $\mathrm{Yb_2Ti_2O_7}$ in a CFM phase would, however, require a drastic revision of the parameters reported for the material, since the exchange parameters pertinent to $\mathrm{Yb_2Ti_2O_7}$ seem to place the compound far away from this new state of matter.

If the QSI phase really exists in $\mathrm{Yb_2Ti_2O_7}$, more investigations on the spin liquid phase above $T_\mathrm{c}$ may provide important evidence. Possible monopole dynamics at temperatures $\sim1.5$ K was recently reported \cite{pan,tokiwa}. Furthermore, the broad anomaly in heat capacity, which we associate here with the inelastic continuum developing at low temperatures, seems to configure an important turning point in the magnetic dynamics of the compound without any obvious consequences in the elastic scattering. We stress that more work, both theoretical and experimental, preferably carried out on well characterised single crystals, is necessary to clarify the blurred physical picture that was built for $\mathrm{Yb_2Ti_2O_7}$ throughout the years.

\section{Conclusions}
In conclusion, we have shown that our $\mathrm{Yb_2Ti_2O_7}$ powder sample adopts a long-range ferromagnetic order below the sharp phase transition at $T_\mathrm{c}=280$ mK. We complement our diffraction data with high resolution inelastic neutron scattering measurements. Powder spherical averaging hides information about directions in reciprocal space that would be fundamental in order to perform a robust quantitative analysis on the $\mathrm{Yb_2Ti_2O_7}$ spectra. Nevertheless, we show that sharp excitation modes clearly exist and that they loosely correspond to what is predicted from high magnetic field data. No other sample of $\mathrm{Yb_2Ti_2O_7}$, powder or single crystal, has been reported to display gapped magnetic excitations in the ordered phase, even though these are expected given the strong exchange anisotropy of the Hamiltonian. In the ferromagnetic state, spin fluctuations translated into a persistent quasielastic scattering are still present above 50 mK. Moreover, a broad continuum develops upon cooling well above $T_\mathrm{c}$. It is difficult to say whether this pyrochlore can be placed in one of the many exotic phases that are theoretically expected to be present in frustrated magnets. We believe that our measurements reveal important new information on the ground state of $\mathrm{Yb_2Ti_2O_7}$ and pave the way for a better understanding of the nature of the ground state of this frustrated magnet.    

\section{Acknowledgements}
We acknowledge insightful discussions with Karen Friese, Owen Benton, Jason S. Gardner, Shigeki Onoda and S\'andor T\'oth. This work would not be possible without the technical support of the sample environment teams of FRM II, Forschungszentrum J\"ulich and ISIS. We gratefully acknowledge Heiner Kolb, Berthold Schimtz and the whole ISIS cryogenics team (with special mention to Chris Lawson) for their invaluable help with the dilution fridges. The work was partially performed at the neutron spallation source SINQ (PSI, Switzerland) and supported by the Science and Technology Facilities Council STFC. V.P.A. was partially supported by CNPq-Brasil.

\bibliography{RevModPhys.82.53,0034-4885-80-1-016502,0034-4885-77-5-056501,PhysRevLett.106.187202,PhysRevX.1.021002,Lieh,PhysRevLett.119.057203,0953-8984-27-16-166002,PhysRevB.92.064425,PhysRevLett.109.097205,PhysRevB.86.174424,0953-8984-13-41-318,PhysRevB.92.134420,PhysRevB.93.100403,PhysRevB.95.094407,0953-8984-28-42-426002,PhysRevLett.115.267208,PhysRevB.95.094422,PhysRevB.89.224419,PhysRevB.93.064406,S092145269901399X,Carvajal199355,SUBRAMANIAN198355,Sala,PhysRevB.95.094431,science,PhysRevLett.88.077204,PhysRevB.70.180404,PhysRevB.84.174442,PhysRevLett.103.227202,Chernyshev,stone,PhysRevLett.110.127207,PhysRevB.87.134408,jps_jpsj72_3014,PhysRevLett.119.127201,PhysRevB.87.205130,PhysRevLett.108.037202,pan,tokiwa,PArmitage}

\end{document}